# Mechanical motion tuned carrier transport characteristic of dynamic diode

*Zunshan Yang[1], Huikai Zhong[2], Can Wang[1], Yanghua Lu[1], Lixuan Feng[1], Xutao Yu[1], Chang Liu[1] and Shisheng Lin*[1,3,4]*

[1]College of Information Science and Electronic Engineering, Zhejiang University, Hangzhou, 310027, P. R. China

[2]Key Laboratory of Nondestructive Test (Ministry of Education), Nanchang Hangkong University, Nanchang 330063, China

[3]State Key Laboratory of Modern Optical Instrumentation, Zhejiang University, Hangzhou, 310027, P. R. China

[4]Key Laboratory of Advanced Micro/Nano Electronic Devices & Smart Systems and Applications of Zhejiang, Zhejiang University, Hangzhou, 310027, P. R. China

*Correspondence: shishenglin@zju.edu.cn.

**Abstract:**

Since the invention of dynamic diode, its physical properties and potential applications have attracted wide attentions. A lot of attempts have been made to harvest the rebounding current and voltage of dynamic diode. However, the underlying physical mechanism of its carrier transport characteristic was rarely explored carefully. Here, the electrical transport properties of the dynamic diode are systematically investigated with a mechanical motion tuned method, where the dynamic current-voltage curve shows a gentler growth trend compared to the static curve. The rebounding current increases with motion velocity and contact force, resulting in a reduced current with the same bias voltage and an oscillation current with a changing velocity and force. Besides, we propose a circuit model with an accurate mathematical formula expression to describe the oscillation current, where an imaginary parameter $\eta_0$ is creatively added to the exponential growth term. This work shows a physical picture of adjust microscopic carrier motion with macroscopic mechanical motion, which provides strong theoretical support for designing dynamic diode devices with better performance in the future.

# 1. Introduction

Owing to the increasing energy requirement of modern society, there has been continuous research for high-efficient and green generators.[1] Meanwhile, dramatic progress has been made in miniaturizing electronics devices over the past decades,[2] which drives the corresponding generator to develop towards miniaturization as a result. Recently, many novel devices have been proposed for in-situ harvesting energy directly from the living environment, such as solar energy,[3] thermoelectric energy [4] and tidal energy,[5] which realize the conversion of photon energy, thermoelectric energy and internal energy to electricity with microscopic interaction mechanism. As a clean and abundant source, the mechanical energy of mutual motion in our daily lives is available around the clock. However, the mechanical energy is usually greatly harvested by electromagnetic effect, [6] triboelectric effect [7] and piezoelectric effect,[8] which are challenging in miniaturization and integration application, besides the output is always alternating current.[9]

Dynamic diode has been widely experimental demonstrated to be capable of generating direct current,[10] which do not require additional rectifier circuits to power electrical devices. Different from the previous power generation mechanism, which mainly involves the photon-electron and charge-electron interaction, dynamic diode shows a physical picture of adjust microscopic carrier motion with macroscopic mechanical motion. In our previous work, a current-density output up of 40.0 A m$^{-2}$ is achieved on a moving van der Waals Schottky Diode and we have successfully converted the mechanical energy into direct current with a PCE over 20%.[11] The dynamic diode is based on the Schottky diode formed by metal/semiconductor structure,[12] where a continuous flow of rebound current is generated through the continuous destruction and establishment of Schottky junction in the dynamic process.[13] The high power density, flexibility and portability property of the dynamic diode generator can meet the new requirement for in-situ sensors from the Internet of

Things and wearable electronic devices.[14] Recently, dynamic diodes have been applied to many different conditions to obtain electricity energy, like harvesting mechanism energy at very low temperatures,[15] wind energy[16] and light energy.[17] However, previous work are mainly focus on increasing power output and proposing new power generation methods, the physical mechanism of dynamic diode still needs to be systematically investigated.

The proposed physical framework mainly concerns the breakdown of the symmetry between the diffusion current and drift current through the interaction between mechanical energy and electrons in semiconductors.[18] However, the mathematical description of the dynamic diode is urgently needed for better designing the output of the dynamic diode for wider applications.[19] In this paper, we try to have a deep insight into the rectification effect of the dynamic diode, which exhibits a switching characteristic that distinguish it from static Schottky junction. The current-voltage (I-V) curve of the dynamic diode can be modulated by the mechanical motion between metal and semiconductor, which provides new understanding of the dynamic diode. Besides, an equivalent circuit model of the dynamic diode is proposed to describe the dynamic carrier transport process of the dynamic diode, and the oscillating I-V curve can be well fitted by the periodical sine function. This work systematically explores the electric transport characteristic of dynamic diode and provides theoretical support for designing dynamic diode devices with better performance in the future.[20]

## 2. Experimental section

The schematic structure of the dynamic diode and experimental setup are illustrated in Figure 1a, where a cooper foil is pressed on a silicon substrate surface to form a Schottky diode. A pressure sensor is connected to the copper foil which can control the force applied to the copper foil surface, while the silicon wafer is fixed to a motor which can rotate at a controlled speed. At this configuration, the copper foil can move with controlled velocity on the silicon substrate while maintaining Schottky contact. The silicon substrate is a single-side polished N-type wafer with a resistivity of 40-50 m·Ω, silver epoxy and an 80 nm thick gold layer are patterned as two electrodes on copper foil and silicon substrate respectively to form good ohmic contact. A Keithley 2400 is connected to the two electrodes using copper wires to record the static and dynamic I-V characteristic curves.

Figure 1b shows the static and dynamic states I-V curves of the dynamic diode, while a constant force of 5.3 N is applied on the copper foil to ensure the Schottky contact. As we can see, the static I-V curve of the device shows a good rectification behavior with a switching voltage of about 1.2 V, indicating the good Schottky junction contact. In the positive bias stage, the current shows a near exponential increasing begin from the switching voltage, which reaches 571 mA when the voltage increases to 5 V. However, in the dynamic state with a moving velocity of 12 cm/s, a similar rectification behavior was also observed with the same switching voltage. However, the dynamic curve shows a relatively slow growth curve in the positive bias stage, where the current only reaches 112 mA when the voltage increases to 5 V, indicating that the dynamic Schottky diode has similar switching characteristics compared to the static Schottky diode. I-V curves of dynamic diodes composed of silicon and other metals have also been tested, as shown in Figure S1, where similar electrical transport characteristic can be observed clearly.

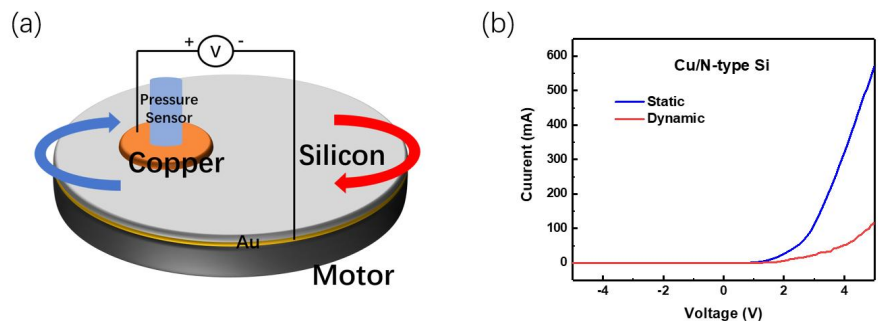

**Figure 1**. (a) Experimental design of dynamic didoes for uniform motion. (b) Current-voltage characteristic curve of Schottky diode in static and dynamic state respectively.

In order to gain a clearer understanding of the underlying physics, more systematic experiments were conducted. First, we checked the dependence of the I-V curve on the moving velocity of the copper foil. As shown in **Figure 2a**, the I-V curves are tested when the copper foil is moving at different velocities on the silicon substrate surface. As we can see, the current in the dynamic Schottky diode decreases as the motion velocity of copper foil increases until 10 cm/s. The I-V curve hardly changes when the speed is greater than 10cm/s. When the copper foil moves at a velocity of 4 cm/s, the current reaches 122 mA as the voltage increases to 4 V, and this value decreases to 51 mA as the velocity increases to 12 cm/s, which suggest greater velocity leads to less current at the same bias voltage. The fluctuation of the dynamic curve can be attributed to the jitter of the rotating motor. At the same time, we conducted the similar experiment on a uniformly varying velocity condition. As we can see from figure 2b , when the velocity changes uniformly with a maximum value of 10 cm/s, the dynamic I-V shows a oscillating behavior at the positive bias voltage region compared to the static I-V curve，which further prove that the I-V curve of the dynamic diode can be adjusted by mutual motion of the two components of the diode.

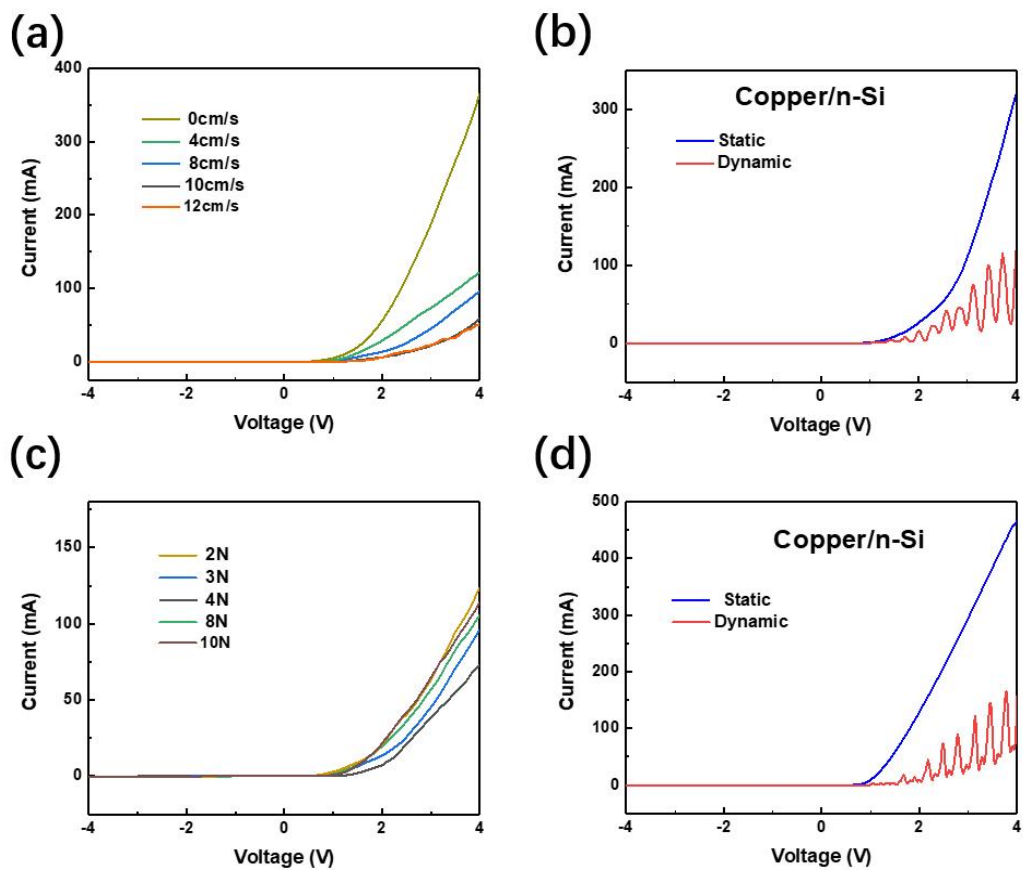

**Figure 2.** (a) I-V curve of dynamic diode in different velocity. (b) Static curve and dynamic curve when the velocity varies uniformly. (c) I-V curves of dynamic diodes under different forces. (d) Static curve and the dynamic curve when a periodic change force is applied.

Besides, we have measured the I-V curve by varying the force applied to the copper foil. The applied force is another important parameter of the Schottky junction. As illustrated in Figure 2c, the conduction characteristics of dynamic diode with different applied forces are tested. It can be observed that as the force applied to the device increases, the current first decreases and reaches a minimum when the applied force is 4 N, and then increases again. In order to find out whether the applied force has a similar modulating effect on the rectification characteristics of dynamic diode, we have checked the dependence of the I-V curve of dynamic diode on the contact force applied to copper foil when the velocity is constant, as shown in Figure 2d. The variation of applied force with time is shown in Figure S2. As we can see, the I-V curve also shows an oscillating behavior at the positive bias voltage region compared to the static I-V curve. To clearly show the difference between the I-V curve in the dynamic state and the static state, we subtract the dynamic current from the static current to obtain the curve shown in Figure S3. The two series of experiments show that the I-V curve of dynamic diode can be tuned with changing motion velocity and contact force, indicating a physical picture of adjust microscopic carrier motion with macroscopic mechanical motion.

## 3. Results and discussion

To further explore the physical mechanism of the rectification characteristics of dynamic diodes within the theoretical framework of semiconductor, the homeostasis establishment and interfacial depletion layer changes under bias and displacement are explored. The work functions of N-type silicon and Cu are about 4.32 eV and 4.5 eV respectively. Thus, the majority carriers of silicon wafer will transfer to copper foil which leads to a diffusion current. Then a depletion layer is formed at the surface of semiconductor. Inside the depletion layer, the space charge is mainly the ionizing donor, and the electron concentration is much smaller than in the semiconductor body, forming a typical barrier layer. And the built-in electric field is directed from the semiconductor body to the surface. The minority carriers drift directionally under the built-in electric field, forming drift current and canceling with the diffusion current. As the metal foil maintains moving, the depletion layer at the boundary of the metal foil is destroyed because of the separation of copper and the silicon. In these parts of semiconductor, however, the built-in electric field does not immediately disappear. After losing the concentration gradient of the interface of copper and silicon, the diffusing majority carriers hit the left built-in electric field and is bounced back and forms a rebound current.

We propose that the much smaller current in the dynamic state is due to the continuous destruction and reconstruction of the depletion layer during the moving process, which break up the carrier distribution.[11] Illustrated in Figure 3a, after losing the concentration gradient at the Cu-Si interface, the diffused carriers in the dynamic Schottky junction hit the built-in electric field on the semiconductor side and are bounced back, forming a rebound current. When the dynamic diode is under positive bias, the barrier height decreased and the built-in electric field is weakened and a bias current from semiconductor towards metal is formed. Since the diffusion and drift currents are still in equilibrium, the current obtained from the I-V test is the net current of the bias and rebound current. As shown in Figure 3b, the bias current is partly offset by the rebound current, which explains why the dynamic curve is lower than the static curve. The bias current is in the same direction as the diffusion current, so the electrons in the bias current can also be rebounded by the built-in electric field. Thus, the difference between the current through the static junction and the dynamic junction increases with increasing bias voltage. And changes in parameters such as speed and applied force cause changes in the trend of I-V curve of the dynamic Schottky junction by affecting the magnitude of the rebound current. As shown in Figure 3c, the voltage and current outputs increase with the moving velocity at low velocity region. When the velocity increases to more than 10 cm/s, the output reaches a saturation state. Figure 3d shows that the dependence of voltage and current outputs on the contact force, where the maximum value is achieved at a 4 N contact force. The data are obtained from the peaks of the rebound current of the dynamic diode, as shown in Figure S4 and S5. Since the rebound current directly affects the net current, the variation in applied force and speed changes the current in the dynamic Schottky diode. This means that we can regulate the rectification characteristic of the dynamic diode by controlling pressure and speed.

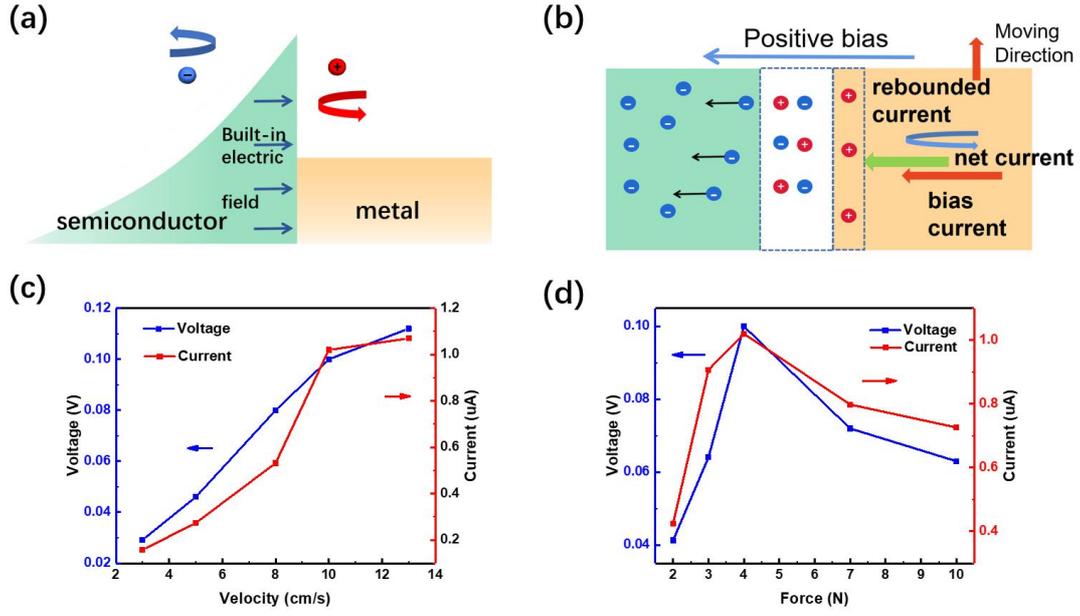

**Figure 3.** (a) The energy band diagram of carrier bounce mechanism. (b) The diagram of carrier distribution and current composition when the device is applied with a voltage bias. The voltage and current outputs of rebound carriers with (c) different speeds and (d) different force applied to the junction.

The I-V curve in static state is fitted and proved to conform to the I-V curve of static diode in the case of series resistance. The current expression of an ideal diode can be expressed as:

$$J = J_s * \left[exp\left(\frac{qV}{K_0 T}\right) - 1\right] \tag{1}$$

$$J_s = \frac{qD_n n_{p0}}{L_n} + \frac{qD_p p_{n0}}{L_p} \tag{2}$$

$J_s$ is saturation current density, q is electron charge, V is the voltage applied, $k_0$ is Boltzmann's constant and T is the temperature. And the result of the fitting of the I-V curve we tested in static state is

$$J = J_s * \left[exp\left(\frac{qV}{K_0 T}\right) - 1\right] + \frac{V}{R} \tag{3}$$

indicating that the heterojunction is a non-ideal diode with a series resistance.

Based on the above conclusions, we propose that the model of dynamic diode in circuit is a diode with series resistance, which has the characteristic of single guide pass. And the slope of its I-V curve varies with the parameter of dynamic diode, such as the force applied to the device and the speed metal film moved. When the rectification characteristic and the parameters such as force and speed are taken into account, then dynamic diode can be equivalent to an electric circuit as shown in

Figure 4a. The systematically research on the rectification characteristic of dynamic diode shows that the device is a diode with a series variable resistance that varies with internal parameters such as force and speed. To verify this model, sets of data have been tested at a uniformly changing speed. According to the circuit model, the I-V curve of dynamic diode can be expressed as

$$J = J_s * [\exp(\eta * V) - 1] + \frac{V}{R} \qquad (4)$$

where the $\eta = \frac{q}{k_0 T} + i * \eta_0$, $\eta_0$ is a coefficient that measures cyclical changes in internal conditions such as pressure and velocity, while the detail of atomic physical meaning of $\eta_0$ calls for more researches. According to Euler's formula, only the real part of the imaginary is considered when fitting. The first term in $\eta$ determines the slope of the I-V curve of dynamic diode, while the second term determines the frequency of oscillation. The $\frac{V}{R}$ is the resistance term of dynamic diode. According to the expression, the experimental data is fitted by Matlab software, as shown in Figure 4c. The fitted curves basically conform to the trend of I-V characteristics, proving our theory.

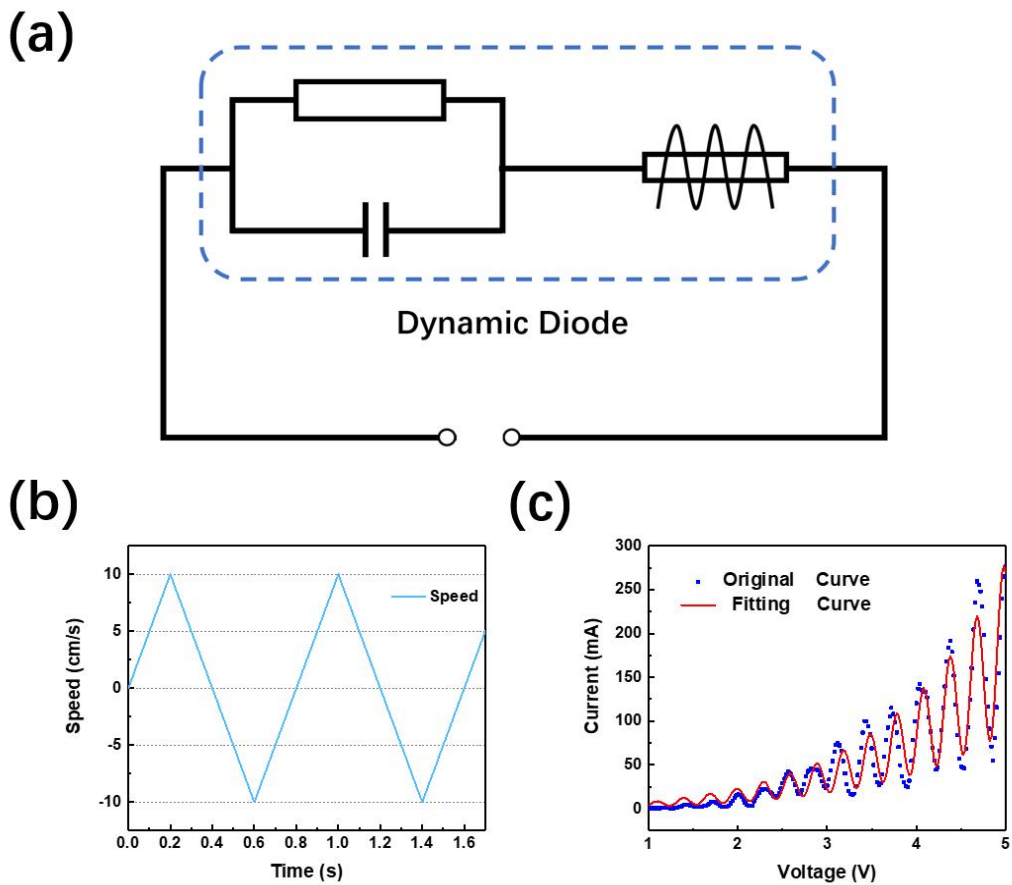

**Figure 4.** (a) Equivalent circuit diagram of dynamic diode. (b) Motion velocity of copper foil when obtaining experimental data. (c) According to the circuit model, I-V curve of typical dynamic diode is fitted by Matlab software.

## 4. Conclusions

In conclusion, the basic physical mechanism of the carrier transport properties of dynamic diodes is carefully investigated with a mechanical motion tuned method. We find that the dynamic current-voltage curve shows a gentler growth trend compared to the static curve, caused by the rebound current. Furthermore, the rebounding current of dynamic diode varies with motion velocity and applied force, resulting in a fluctuating current curve. Based on this mechanism, the electrical characteristic of dynamic diode can be modified by motion velocity and contact force, which means a novel diode device different from the traditional static diode. In addition, the applied bias voltage can amplify the effect of the rebound current by changing the barrier height and affecting the built-in electric field. Finally, we propose an equivalent circuit model for the dynamic diode to help describe the dynamic carrier transport process. This work provides a deep insight into the electrical properties of dynamic diode, proving a strong support for the rebound carrier theory and laying a solid theoretical foundation for the future application.

## Acknowledgements

S.S.L. thanks the support from the National Natural Science Foundation of China (Grant Nos. 51202216, 51551203, and 61774135), Special Foundation of Young Professor of Zhejiang University (Grant No. 2013QNA5007), and Outstanding Youth Fund of Zhejiang Natural Science Foundation of China (Grant No. LR21F040001). Y.H.L. thanks the support from the China Postdoctoral Science Foundation (Grant No. 2021M692767). H.K. Zhong thanks the Open Foundation of Key Laboratory of Nondestructive Testing Technology (Ministry of Education, Nanchang Hangkong University, Grant No. EW202208373).

# Conflict of Interest

The authors declare no conflict of interest.

# Author Contributions

S.L. designed the experiments, analyzed the data, and conceived the study. Z.Y. and C.W. designed and carried out the experiments, discussed the results, and wrote the paper. Y.L., L.F., X.Y., C.L., Y. Y., R. S. discussed the results and assisted with experiments. All authors contributed to the preparation of the manuscript.

# Data Availability Statement

The data that support the findings of this study are available from the corresponding author upon reasonable request.

# References

[1]  a)J. Meng, Z. H. Guo, C. Pan, L. Wang, C. Chang, L. Li, X. Pu, Z. L. Wang, ACS Energy Letters 2021, 6, 2442; b)Zhong, Huikai, Wang, Fengchao, Chen, Hongsheng, Hengan, Shisheng; c)W. Xuan, L. Zhi, K. Müllen, Nano Letters 2008, 8, 323.
[2]  a)X. Gou, H. Xiao, S. Yang, Applied Energy 2010, 87, 3131; b)Q. Zhang, Y. Sun, W. Xu, D. Zhu, Advanced Materials 2014, 26, 6829; c)F. R. Fan, W. Tang, Z. L. Wang, Advanced Materials 2016, 28, 4283.
[3]  a)B. Tian, X. Zheng, T. J. Kempa, Y. Fang, N. Yu, G. Yu, J. Huang, C. M. Lieber, Nature 2007, 449, 885; b)Y. Qin, X. Wang, Z. L. Wang, Nature 2008, 451, 809.
[4]  Y. Yang, W. Guo, K. C. Pradel, G. Zhu, Y. Zhou, Y. Zhang, Y. Hu, L. Lin, Z. L. Wang, Nano Letters 2012, 12, 2833.
[5]  W. Xie, L. Gao, L. Wu, X. Chen, F. Wang, D. Tong, J. Zhang, J. Lan, X. He, X. Mu, Y. Yang, Research 2021, 2021, 5963293.
[6]  S. P. Beeby, R. N. Torah, M. J. Tudor, P. Glynne-Jones, T. O'Donnell, C. R. Saha, S. Roy, Journal of Micromechanics and Microengineering 2007, 17, 1257.
[7]  F.-R. Fan, Z.-Q. Tian, Z. Lin Wang, Nano Energy 2012, 1, 328.
[8]  Z. L. Wang, J. Song, American Association for the Advancement of Science 2006.

[9] a)P. Wang, L. Pan, J. Wang, M. Xu, G. Dai, H. Zou, K. Dong, Z. L. Wang, ACS Nano 2018, 12, 9433; b)C. Zhang, W. Tang, C. Han, F. Fan, Z. L. Wang, Advanced Materials 2014, 26, 3580.

[10] a)Y. Lu, P. Zhang, S. Lin, S. Feng, R. Shen, Y. Xu, Z. Hao, Y. Yan, H. Zheng, X. Yu, 2020; b)Y. Lu, Z. Hao, S. Feng, R. Shen, Y. Yan, S. Lin, iScience 2019, 22, 58; c)S. Lin, R. Shen, T. Yao, Y. Lu, S. Feng, Z. Hao, H. Zheng, Y. Yan, E. Li, Advanced Science 2019, 6; d)M. Zheng, S. Lin, L. Xu, L. Zhu, Z. L. Wang, Advanced Materials 2020, 32, 2000928.

[11] S. Lin, Y. Lu, S. Feng, Z. Hao, Y. Yan, Advanced Materials 2019, 31, 1804398.

[12] J. Bardeen, Physical Review 1947, 71, 717.

[13] a)S. Lin, R. Shen, T. Yao, Y. Lu, S. Feng, Z. Hao, H. Zheng, Y. Yan, E. Li, Advanced Science 2019, 6, 1901925; b)J. Liu, A. Goswami, K. Jiang, F. Khan, S. Kim, R. McGee, Z. Li, Z. Hu, J. Lee, T. Thundat, Nature Nanotechnology 2018, 13, 112.

[14] a)Z. L. Wang, J. Chen, L. Lin, Energy & Environmental Science 2015, 8, 2250; b)Q. Shi, B. Dong, T. He, Z. Sun, J. Zhu, Z. Zhang, C. Lee, InfoMat 2020, 2, 1131.

[15] H. Zheng, R. Shen, H. Zhong, Y. Lu, X. Yu, S. Lin, Advanced Functional Materials 2021, 31, 2105325.

[16] X. Yu, H. Zheng, Y. Lu, R. Shen, Y. Yan, Z. Hao, Y. Yang, S. Lin, RSC Advances 2021, 11, 19106.

[17] Z. Hao, T. Jiang, Y. Lu, S. Feng, R. Shen, T. Yao, Y. Yan, Y. Yang, Y. Lu, S. Lin, Matter 2019, 1, 639.

[18] S. Lin, Y. Lu, J. Xu, S. Feng, J. Li, Nano Energy 2017, 40, 122.

[19] a)J. Liu, M. Miao, K. Jiang, F. Khan, A. Goswami, R. McGee, Z. Li, L. Nguyen, Z. Hu, J. Lee, K. Cadien, T. Thundat, Nano Energy 2018, 48, 320; b)J. Liu, Y. Zhang, J. Chen, R. Bao, K. Jiang, F. Khan, A. Goswami, Z. Li, F. Liu, K. Feng, J. Luo, T. Thundat, Matter 2019, 1, 650.

[20] P. Matyba, K. Maturova, M. Kemerink, N. D. Robinson, L. Edman, Nature Materials 2009, 8, 672.